# Scheme for the implementation of optimal cloning of two pairs of orthogonal states

Zhuo-Liang Cao[*], Wei Song[†]

Department of Physics, Anhui University, Hefei, 230039, P. R. of China

**Abstract**

We propose a feasible scheme to implement the $1 \to 2$ optimal cloning transformation for two pairs of orthogonal states of two-dimensional quantum systems in the context of cavity QED. The copied qubits are shown to be inseparable by using Peres-Horodecki criterion.

[*] Electronic address: zlcao@mars.ahu.edu.cn
[†] Electronic address: wsong@mars.ahu.edu.cn

## 1. Introduction

Quantum information and quantum computing have been attracting a great deal of interesting. They differ in many aspects from the classical theories. One of the most fundamental difference between classical and quantum information is that while classical information can be copied perfectly, quantum information cannot. In particular, it follows from the no-cloning theorem [1] that one cannot create a perfect duplicate of an arbitrary qubit. The no-cloning theorem for pure states is also extended to the case that a general mixed state cannot be broadcast [2]. However, the no-cloning theorem does not forbid imperfect cloning, and the imperfect cloning of quantum states have attracted much attention with the development of quantum information theory.

The imperfect cloning may be divided into two main categories: state-independent and state-dependent. Buzek and Hillery firstly proposed a universal quantum cloning machine(UQCM) [3] for an arbitrary pure state where the copying process is input-state independent, which means this machine does not need any information about the state to be cloned. They use Hilbert-Schmidt norm to quantify distances between the input density operator and the output density operators. The Buzek- Hillery cloning machine has been optimized and generalized in Refs. [4-8]. There is another kind of cloning machine named state-dependent cloning machine. It needs some information about the cloning state, and it was first investigated by Bruss et al. [5] and has been solved completely when the state set contains only two states. Bounds of the global fidelity for state-dependent quantum cloning were derived when the state set contains more than two states [9]. Quantum cloning machines for equatorial qubits have been studied in Refs.[5,10,11], where the fidelity is optimized for states lying on a great-circle of the Bloch sphere. And optimal cloning for two pairs of orthogonal states have also been obtained by Bruss et al. [12]. Another state-dependent quantum cloning machine was introduced by Duan and Guo [13,14]. They found that the states could be cloned perfectly with some probability less than 1, when the states are linearly independent.

On the other hand, it is important to obtain a physical means to carry out the different cloning process, and quantum networks for universal cloning have been proposed by Buzek et al. [15], the networks are also constructed for some kinds of state-dependent cloning machine [16]. Recently, several schemes for realization of universal quantum cloning machine have been suggested with quantum optics[17] and cavity QED[18], and experiments have been performed with linear optics[19], parametric down conversion[20], and NMR[21].

In this paper, using the approach presented in Ref.[15], we show that the optimal cloning for two pairs of orthogonal can be realized by networks consisted of quantum rotation gates and controlled NOT gates. Then we propose a scheme to realize the networks within cavity QED techniques. We further analyze the inseparability of the output qubits.

## 2. Networks of optimal cloning of two pairs of orthogonal states

We consider an ensemble of input states that consists of two pairs of orthogonal states for a two-dimensional quantum system. These four states can be parametrized in the Bloch sphere representation with a single parameter. The four Bloch vectors $\hat{m}_i$ for the states $|\Psi_i\rangle$ with

$$|\Psi_i\rangle\langle\Psi_i| = \frac{1}{2}(I + \hat{m}_i \cdot \hat{\sigma}) \qquad i = 1,\cdots,4 \tag{1}$$

where $I$ is the identity operator and $\sigma_i$ with $i = x, y, z$ is the Pauli matrix, $\hat{m}_i$ is given by

$$\hat{m}_1 = \begin{pmatrix} \sin\varphi \\ 0 \\ \cos\varphi \end{pmatrix}, \quad \hat{m}_2 = \begin{pmatrix} -\sin\varphi \\ 0 \\ \cos\varphi \end{pmatrix}, \quad \hat{m}_3 = \begin{pmatrix} -\sin\varphi \\ 0 \\ -\cos\varphi \end{pmatrix}, \quad \hat{m}_4 = \begin{pmatrix} \sin\varphi \\ 0 \\ -\cos\varphi \end{pmatrix} \tag{2}$$

In this representation the four vectors are lying in the $x,z$-plane, the two pairs of orthogonal states are given by $\{|\Psi_1\rangle, |\Psi_3\rangle\}$ and $\{|\Psi_2\rangle, |\Psi_4\rangle\}$. We could parametrize the states $|\Psi_i\rangle$ with the real parameters $\alpha$ and $\beta$ with $\alpha^2 + \beta^2 = 1$.

$$|\Psi_1\rangle = \alpha|0\rangle + \beta|1\rangle, \ |\Psi_2\rangle = \alpha|0\rangle - \beta|1\rangle, \ |\Psi_3\rangle = \beta|0\rangle - \alpha|1\rangle, \ |\Psi_4\rangle = \beta|0\rangle + \alpha|1\rangle \tag{3}$$

The case of optimal cloning for these two pairs of orthogonal states has already been found by Bruss et al. in Ref. [12]. They proposed the following cloning transformation for the input (3).

$$|0\rangle_{a_1}|0\rangle_{a_2}|X\rangle \to a|000\rangle_{a_1a_2a_3} + b(|011\rangle_{a_1a_2a_3} + |101\rangle_{a_1a_2a_3}) + c|110\rangle_{a_1a_2a_3} \tag{4a}$$

$$|1\rangle_{a_1}|0\rangle_{a_2}|X\rangle \to a|111\rangle_{a_1a_2a_3} + b(|100\rangle_{a_1a_2a_3} + |010\rangle_{a_1a_2a_3}) + c|001\rangle_{a_1a_2a_3} \tag{4b}$$

where $|X\rangle$ is an arbitrary auxiliary state, and

$$a = \frac{1}{2}\left(1+\cos^2\varphi\sqrt{\frac{1}{\sin^4\varphi+\cos^4\varphi}}\right), \quad b = \frac{1}{2}\sin^2\varphi\sqrt{\frac{1}{\sin^4\varphi+\cos^4\varphi}}$$

$$c = \frac{1}{2}\left(1-\cos^2\varphi\sqrt{\frac{1}{\sin^4\varphi+\cos^4\varphi}}\right) \tag{5}$$

Now, following the method proposed by Buzek et al.[15], we show that the quantum cloning transformations for two pairs of orthogonal states can be realized by networks consisting of quantum logic gates. Let us first introduce the method proposed by Buzek et al.[15]. The network is constructed by one- and two-qubit gates. The one-qubit gate is a single qubit rotation operator $\hat{R}_j(\vartheta)$, defined as

$$\hat{R}_j(\vartheta)|0\rangle_j = \cos\vartheta|0\rangle_j + \sin\vartheta|1\rangle_j, \qquad \hat{R}_j(\vartheta)|1\rangle_j = -\sin\vartheta|0\rangle_j + \cos\vartheta|1\rangle_j \tag{6}$$

The two-qubit gate is the controlled NOT gate represented by the unitary matrix

$$\hat{P} = \begin{pmatrix} 1 & 0 & 0 & 0 \\ 0 & 1 & 0 & 0 \\ 0 & 0 & 0 & 1 \\ 0 & 0 & 1 & 0 \end{pmatrix} \tag{7}$$

The controlled NOT gate $\hat{P}_{kl}$ acts on the basis vectors of the two qubits as follows:

$$\hat{P}_{kl}|0\rangle_k|0\rangle_l = |0\rangle_k|0\rangle_l, \quad \hat{P}_{kl}|0\rangle_k|1\rangle_l = |0\rangle_k|1\rangle_l, \quad \hat{P}_{kl}|1\rangle_k|0\rangle_l = |1\rangle_k|1\rangle_l, \quad \hat{P}_{kl}|1\rangle_k|1\rangle_l = |1\rangle_k|0\rangle_l \tag{8}$$

Due to the method proposed by Buzek et al, the action of the copier is expressed as a sequence of two unitary transformations

$$|\Psi^{(in)}\rangle_{a_1}|0\rangle_{a_2}|0\rangle_{a_3} \to |\Psi\rangle^{(in)}_{a_1}|\Psi\rangle^{(prep)}_{a_2 a_3} \to |\Psi\rangle^{(out)}_{a_1 a_2 a_3} \tag{9}$$

The preparation state is constructed as

$$|\Psi\rangle^{(prep)}_{a_2 a_3} = \hat{R}_2(\vartheta_3)\hat{P}_{32}\hat{R}_3(\vartheta_2)\hat{P}_{23}\hat{R}_2(\vartheta_1)|0\rangle_{a_2}|0\rangle_{a_3} \tag{10}$$

In the case of cloning transformation of two pairs of orthogonal states, the preparation state may be written as

$$|\Psi\rangle^{(prep)}_{a_2 a_3} = a|00\rangle_{a_2 a_3} + b\left(|01\rangle_{a_2 a_3} + |10\rangle_{a_2 a_3}\right) + c|11\rangle_{a_2 a_3} \tag{11}$$

The solution for $\vartheta_1, \vartheta_2,$ and $\vartheta_3$ turns out to be

$$\vartheta_2 = \arcsin\left[\frac{\sqrt{2}}{2}(a+c)\right] - \frac{\pi}{4} = 0, \vartheta_1 = \vartheta_3 = \frac{1}{2}\arcsin\left[\frac{2b}{\cos\vartheta_2 - \sin\vartheta_2}\right] = \frac{1}{2}\arcsin(2b) \quad (12)$$

Once the qubits of the quantum copier are properly prepared then the copying of the initial state $|\Psi\rangle_{a_1}^{(in)}$ of the original qubit can be performed by sequence of three controlled NOT operations (see fig.1)

$$|\Psi\rangle_{a_1 a_2 a_3}^{(out)} = \hat{P}_{a_2 a_3} \hat{P}_{a_3 a_1} \hat{P}_{a_1 a_2} |\Psi\rangle_{a_1}^{(in)} |\Psi\rangle_{a_2 a_3}^{(prep)} \quad (13)$$

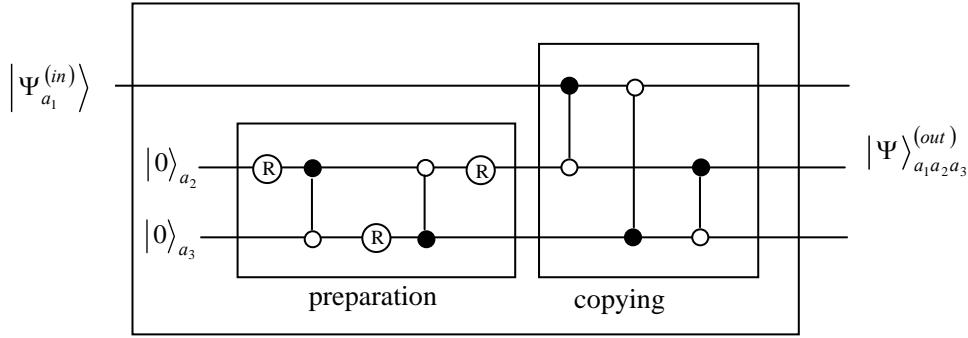

FIG.1. The networks of optimal cloning for two pairs of orthogonal states. The logical controlled NOT $\hat{P}_{kl}$ given by Eq.(8) has as its input a control qubit(denoted as ●) and a target qubit (denoted as ○). The action of the single-qubit operator R is specified by the transformation (6)

### 3. Implementation of optimal cloning of two pairs of orthogonal states

Next, we will propose a feasible scheme for this cloning process of a qubit represented by a superposition of the two long-living circular states $|g\rangle$ and $|e\rangle$ of a Rydbery atom. The cloning process is accomplished by three unitary operations and five controlled NOT gates which we have depicted in the above networks. The single qubit unitary operation used in the scheme is implemented by the resonant interaction of the Rydberg atom with a classical field [22]. Assume that a classical field is resonant with the $|i\rangle \leftrightarrow |j\rangle$ ( $i,j = f, g, e; i \neq j$ ) transition, it can be easily shown that the evolution operator of the atomic state under the basis $\{|i\rangle, |j\rangle\}$ is

$$U_{ij}(\theta,\omega) = \begin{pmatrix} \cos\theta & -ie^{-i\omega}\sin\theta \\ -ie^{i\omega}\sin\theta & \cos\theta \end{pmatrix} \quad (14)$$

where $\theta = \Omega_{ij}|\alpha|t$, $\alpha = |\alpha|e^{i\omega}$ is the complex amplitude of the classical field, $\Omega_{ij}$ is the atom-field coupling constant, and the state $|k\rangle(k = f, g, e; k \neq i, j)$ take no effect during this stage, by choosing $\omega = \frac{3}{2}\pi$, we can get the transformation of Eq. (6). The controlled not gates have been realized with cavity QED techniques in Ref. [23]. The whole experiment process is depicted in Fig.2.

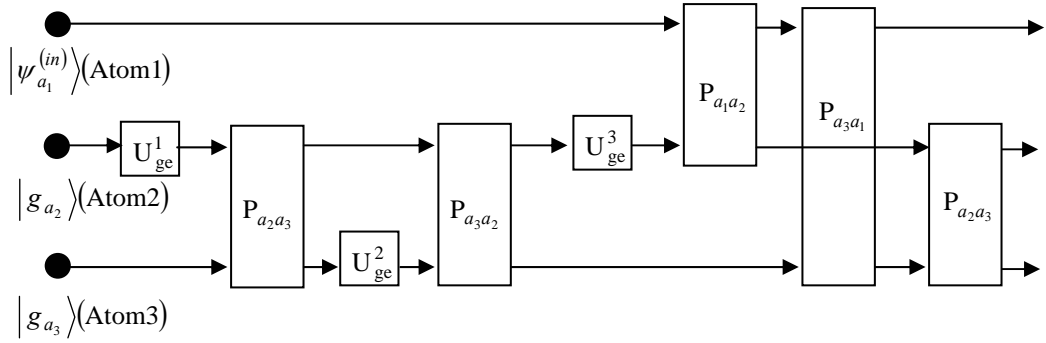

FIG.2. This is the schematic diagram of the setup which could implement the optimal cloning for two pairs of orthogonal states. $|\psi_{a_1}^{(in)}\rangle$ is the input qubit prepared in a superposition of Eq.(3). $U_{ge}^n (n = 1, 2, 3)$ can be realized by the resonant interaction of Rydberg atom with a classical field, which is specified by the transformation(14).The controlled NOT gate between two atoms has been discussed in detail within cavity QED techniques in Ref. [23]

In the following let us briefly discuss the fidelity of this scheme, the fidelity of this cloning process has been discussed in Ref. [12], we will only discuss the decrease of the fidelity induced by the errors. In our scheme the errors may be induced by the single qubit rotation operation and the C-NOT gate, and the C-NOT gate will not affect the fidelity of this experiment, so we will mainly discuss the errors which result from the single qubit rotation operation. We define the ideal output state is $|\Psi\rangle_{ideal}$, the actual output state is $|\Psi\rangle_{actual}$, so the fidelity is given by $F = |_{ideal}\langle\Psi|\Psi\rangle_{actual}|^2$. Suppose there are small mistakes during the single qubit rotation operation: $\vartheta_1' = \vartheta_1 + \Delta\vartheta_1$, $\vartheta_2' = \vartheta_2 + \Delta\vartheta_2$, $\vartheta_3' = \vartheta_3 + \Delta\vartheta_3$, where $\Delta\vartheta_1, \Delta\vartheta_2, \Delta\vartheta_3$ are the errors of the three angles respectively. For simplicity, we could assume $\Delta\vartheta_1 = \Delta\vartheta_2 = \Delta\vartheta_3 = \Delta\vartheta$, when the value of $\Delta\vartheta$ is small, we have $\cos(\vartheta + \Delta\vartheta) = \cos\vartheta - \Delta\vartheta\sin\vartheta$, $\sin(\vartheta + \Delta\vartheta) = \sin\vartheta + \Delta\vartheta\cos\vartheta$, then $F$ can be simplified as

$$F(\varphi, \Delta\vartheta) = \frac{1}{N^2}\left\{a\left[a - 2b\Delta\vartheta + \frac{1}{2}\left(1 - \sqrt{1-4b^2}\right)\Delta\vartheta\right] + 2b\left(b - b\Delta\vartheta + \sqrt{1-4b^2}\Delta\vartheta\right)\right.$$

$$\left. + c\left[c + 2b\Delta\vartheta + \frac{1}{2}\left(1 + \sqrt{1-4b^2}\right)\Delta\vartheta\right]\right\}^2 \tag{15}$$

where N is the normalized factor and it is given by

$$N^2 = \left[a - 2b\Delta\vartheta + \frac{1}{2}\left(1 - \sqrt{1-4b^2}\right)\Delta\vartheta\right]^2 + 2\left(b - b\Delta\vartheta + \sqrt{1-4b^2}\Delta\vartheta\right)^2$$

$$+ \left[c + 2b\Delta\vartheta + \frac{1}{2}\left(1 + \sqrt{1-4b^2}\right)\Delta\vartheta\right]^2 \tag{16}$$

Noting the value of $\Delta\vartheta$ is small, then we only need to consider the first order of $\Delta\vartheta$, after little algebra we still have $F = 1$, so we conclude when the error of the three angles during rotation operation is small, we can still get a high fidelity output state. The single qubit operation used in the scheme can be easily implemented by the resonant interaction of the Rydberg atom with a classical field [22], once the C-NOT gate based on cavity QED techniques could be implemented in experiment our scheme might be realized.

Finally, we would like to determine whether the two copiers are quantum-mechanically entangled. To do so, we can utilize the Peres-Horodecki theorem [24,25], which provides us with a simple algorithm for determining whether or not a general two-qubit state is entangled. All that is necessary is to calculate the eigenvalues of the partial transpose of the state's density matrix. According to the theorem, a two-qubit state is entangled if and only if at least one of these eigenvalues is negative. In this case we assume the copiers appear in the $a_2$ and $a_3$ qubits, without loss of generality, we will discuss the case that the original qubit is in the superposition state $|\Psi_1\rangle^{(in)}_{a_1} = \alpha|0\rangle + \beta|1\rangle$. The reduced density matrix of the two qubits in the $\{|00\rangle_{a_2 a_3}, |01\rangle_{a_2 a_3}, |10\rangle_{a_2 a_3}, |11\rangle_{a_2 a_3}\}$ basis is

$$\hat{\rho}^{(out)}_{a_2a_3} = \begin{pmatrix} \alpha^2a^2+\beta^2b^2 & \alpha\beta(ac+b^2) & \alpha\beta(ab+bc) & ab \\ \alpha\beta(ac+b^2) & \alpha^2b^2+\beta^2c^2 & bc & \alpha\beta(ab+bc) \\ \alpha\beta(ab+bc) & bc & \alpha^2c^2+\beta^2b^2 & \alpha\beta(ac+b^2) \\ ab & \alpha\beta(ab+bc) & \alpha\beta(ac+b^2) & \alpha^2b^2+\beta^2a^2 \end{pmatrix} \quad (17)$$

The partial transpose of Eq.(17) is

$$\hat{\rho}^{T_2}_{a_2a_3} = \begin{pmatrix} \alpha^2a^2+\beta^2b^2 & \alpha\beta(ac+b^2) & \alpha\beta(ab+bc) & bc \\ \alpha\beta(ac+b^2) & \alpha^2b^2+\beta^2c^2 & ab & \alpha\beta(ab+bc) \\ \alpha\beta(ab+bc) & ab & \alpha^2c^2+\beta^2b^2 & \alpha\beta(ac+b^2) \\ bc & \alpha\beta(ab+bc) & \alpha\beta(ac+b^2) & \alpha^2b^2+\beta^2a^2 \end{pmatrix} \quad (18)$$

We have the following four eignvalues:

$$\left\{ \frac{a^2+b^2 \pm \sqrt{(a^2-b^2)^2+4b^2c^2}}{2}, \quad \frac{b^2+c^2 \pm \sqrt{(b^2-c^2)^2+4a^2b^2}}{2} \right\} \quad (19)$$

Noting the condition $c^2 < a^2$, we see that the value of $\frac{b^2+c^2-\sqrt{(b^2-c^2)^2+4a^2b^2}}{2}$ is negative, when the original qubit is in the other three superposition states, we have the same eignvalues as above, it follows that the two qubits at the output of the quantum copier are nonclassically entangled.

4. **Conclusion**

There still leaves a number of open questions which we briefly discuss now. Perhaps the most obvious is how to find the $1 \to N$ optimal transformation for two pairs of orthogonal states. Another generalization would be the optimal cloning transformation for *d* pairs of orthogonal states in *d*-dimensional space. It is also important to find other feasible experimental schemes to realize the process of optimal cloning of two pairs of orthogonal states which may be linear optics or the other techniques.

In summary, we have proposed the network consisting of quantum gates for optimal cloning of two pairs of orthogonal states, which can be realized within cavity QED techniques. The inseparability of the output qubits has also been shown. This feature is important that we must keep in mind when determining how to make use of the copies.

**Acknowledgement**

This work is supported by Anhui Provincial Natural Science Foundation under Grant No:

03042401, the Key Programof the Education Department of Anhui Province under Grant No: 2004kj005zd and the Talent Foundation of Anhui University.